\documentclass{emulateapj}

\usepackage{xspace}
\usepackage{graphicx}
\usepackage{rotating}
\usepackage{natbib}
\usepackage{apjfonts}
\def\errtwo#1#2#3{$#1^{+#2}_{-#3}$}

\newcommand\zth{$0^\mathrm{th}$}
\newcommand\fst{$1^\mathrm{st}$}

\newcommand\chandra{\textsl{Chandra}\xspace}
\newcommand\nustar{\textsl{NuSTAR}\xspace}

\newcommand\heg{{HEG}\xspace}

\newcommand\heasoft{{HEASOFT}\xspace}

\newcommand\integral{\textsl{INTEGRAL}\xspace}

\newcommand\meg{{MEG}\xspace}

\newcommand\mysou{IGR~J17454$-$2919}

\newcommand\osa{{\tt OSA}\xspace}

\newcommand\swift{\textsl{Swift}\xspace}

\newcommand\xmm{\textsl{XMM-Newton}\xspace}

\newcommand\xselect{{\tt XSELECT}\xspace}
\newcommand\aproxgt{\mathrel{%
     \rlap{\raise 0.511ex \hbox{$>$}}{\lower 0.511ex \hbox{$\sim$}}}}
\newcommand\aproxlt{\mathrel{%
     \rlap{\raise 0.511ex \hbox{$<$}}{\lower 0.511ex \hbox{$\sim$}}}}

\slugcomment{Submitted }

\shorttitle{IGR~J17454$-$2919}
\shortauthors{Paizis et al. 2015}

\begin{document}

\title{Investigating the nature of \mysou~using X--ray and Near-Infrared observations }

\author{A. Paizis\altaffilmark{1}, 
M.~A. Nowak\altaffilmark{2},
J. Rodriguez\altaffilmark{3},
A. Segreto\altaffilmark{4},
S. Chaty\altaffilmark{3,5},
A. Rau\altaffilmark{6},
J. Chenevez\altaffilmark{7},
M. Del Santo\altaffilmark{4},\\
J. Greiner\altaffilmark{6},
S. Schmidl\altaffilmark{8}
}

\altaffiltext{1}{Istituto Nazionale di Astrofisica, INAF-IASF, Via Bassini 15, 20133 Milano, Italy; ada@iasf-milano.inaf.it}
\altaffiltext{2}{Massachusetts Institute of Technology, Kavli Institute for Astrophysics, Cambridge, MA 02139, USA; mnowak@space.mit.edu}
\altaffiltext{3}{AIM  - Astrophysique, Instrumentation et Mod\'elisation (UMR-E 9005 CEA/DSM-CNRS-Universit\'e Paris Diderot) Irfu/Service d'Astrophysique, Centre de Saclay FR-91191 Gif-sur-Yvette Cedex, France}
\altaffiltext{4}{Istituto Nazionale di Astrofisica, IASF Palermo, Via U. La Malfa 153, I-90146 Palermo, Italy }
\altaffiltext{5}{Institut Universitaire de France, 103, boulevard Saint-Michel, 75005 Paris}
\altaffiltext{6}{Max-Planck-Institute for Extraterrestrial Physics, 85741 Garching, Germany}
\altaffiltext{7}{National Space Institute, Technical University of Denmark, Elektrovej 327-328, 2800 Kgs Lyngby, Denmark}
\altaffiltext{8}{Th\"{u}ringer Landessternwarte Tautenburg, Sternwarte 5, 07778 Tautenburg, Germany}

\begin{abstract}
\mysou~is a hard X-ray transient discovered by \integral~on 2014 September 27. 
We report on our 20\,ks \chandra~observation of the source, performed about five weeks after the discovery, as well as on \integral~and \swift~monitoring long-term observations. X--ray broad-band spectra of the source are compatible with an absorbed power-law, $\Gamma\sim$1.6-1.8, ${\rm N_H}\sim$(10--12)$\times 10^{22}\,{\rm cm}^{-2}$, with no trace of a cut-off in the data up to about 100\,keV, and  with an average absorbed 0.5--100\,keV flux of about (7.1--9.7)$\mathrm{\times 10^{-10}~erg~cm^{-2}~s^{-1}}$. 
 With \chandra, we  determine the most accurate  X-ray position of \mysou, $\alpha_\mathrm{J2000}$=17$^\mathrm{h}$ 45$^\mathrm{m}$ 27$^\mathrm{s}$.69, \mbox{$\delta_\mathrm{J2000}$= $-$29$^{\circ}$ 19$^{\prime}$ 53$^{\prime \prime}$}.8 (90\% uncertainty of 0$^{\prime\prime}$.6), 
consistent with the NIR source 2MASS~J17452768--2919534. 
We also include NIR investigations from our observations of the source field on 2014 October 6 with GROND.
With the multi-wavelength information at hand, we discuss the possible nature of \mysou.

\end{abstract}


\section{Introduction}\label{sec:intro}
\setcounter{footnote}{0}

The bulge of our Galaxy contains a variety of high--energy transient and persistent sources. It is a unique environment where we can study a wide range of X--ray intensities, down to the fainter levels. Quantifying the spatial distribution, activity and properties of these sources is essential for population studies and hence for understanding the evolution of our own Galaxy. Large field-of-view instruments with high sensitivity in the hard X--ray energy band, less contaminated by the numerous soft X--ray
emitting stars, are essential ingredients for such a study. In this respect, the \integral~observatory \citep{winkler03, winkler11}, with its regular monitoring and deep observations of the Galactic center, has offered a fundamental contribution \cite[e.g.][]{kuulkers07, revnivtsev08, bird10,  krivonos12, lutovinov13}.

On 2014 September 27, \integral~discovered the new hard X-ray transient source \mysou~\citep{chenevez14}. The source, less than 24$^{\prime}$ from the Galactic Center, was detected by both JEM--X detectors \citep{lund03} in mosaic images obtained from  observations during \integral~revolution 1460 (2014 September 27-30). The  reported JEM--X average fluxes were 6.5$\pm$1\,mCrab (3--10\,keV) and 8.2$\pm$1.7\,mCrab (10--25\,keV).
The source was not detected in previous observations of the region, with a 5$\sigma$ 3--25\,keV upper limit of about 1\,mCrab. \citet{chenevez14} also report on a \swift~2\,ks follow-up observation on 2014 October 2. \mysou~was clearly visible,  $\alpha_\mathrm{J2000}$=17$^\mathrm{h}$ 45$^\mathrm{m}$ 28$^\mathrm{s}$, 
\mbox{$\delta_\mathrm{J2000}$= $-$29$^{\circ}$ 19$^{\prime}$ 55$^{\prime \prime}$} (90\% error confidence of 5$^{\prime \prime}$), only 10$^{\prime \prime}$ from the JEM--X position.

Later on, \citet{chenevez14b} reported that the source flux, observed by \integral/JEM--X on 2014 October 18-20, had increased by about a factor of two with respect to the previous observations, reaching a flux of 10$\pm$1\,mCrab in 3--10\,keV and 15$\pm$2\,mCrab in 10--25\,keV.

On 2014 October 10, \nustar~\citep{harrison13} observed \mysou~for a total of about 29\,ks \citep{tendulkar14}. The energy spectrum could be well described by an absorbed  power-law (${\rm N_H} =3.3 \pm0.6 \times 10^{22}\,{\rm cm}^{-2}$, $\Gamma = 1.46\pm0.06$) with an exponential cut-off ($>$100\,keV) and a broad, asymmetric, iron emission line. The unabsorbed 3--79\,keV flux was 3.96$\mathrm{\times  10^{-10}~erg~cm^{-2}~s^{-1}}$, corresponding to an isotropic luminosity of 3$\mathrm{\times 10^{36}~erg~s^{-1}}$ at 8\,kpc. Similarly to JEM--X \citep{chenevez14}, the 3--79\,keV light curve did not show any evidence for bursts or pulsations. \citet{tendulkar14} noted a 14\% drop in the count rate over the course of their observation, with 25$\pm$3\% rms variability in the form of a power density spectrum (PDS) described by a zero frequency centered Lorentzian with 2\,Hz width and peaking at 1\,Hz.  Additional power in the PDS was seen at low frequencies in the form of a power-law.
According to the authors, the hard power-law index, high energy cut-off, and level of variability are consistent with \mysou~being an accreting black hole (BH) in the hard state, though the possibility of a low magnetic field neutron star (NS) cannot be ruled out.

On 2014 November 3, we observed \mysou~with \chandra/HETGS for  20\,ks. Our \chandra~based position was reported in \citet{paizis15} as $\alpha_\mathrm{J2000}$=17$^\mathrm{h}$ 45$^\mathrm{m}$ 27$^\mathrm{s}$.69, 
\mbox{$\delta_\mathrm{J2000}$= $-$29$^{\circ}$ 19$^{\prime}$ 53$^{\prime \prime}$}.8
(90\% uncertainty of 0$^{\prime\prime}$.6). This position (2.4$^{\prime \prime}$ away from the \swift/XRT one) is consistent with the near-infrared (NIR) source 2MASS~J17452768--2919534. \\
On 2015 February 16--17, \integral~observed again the Galactic center and \citet{boissay15} reported the non detection of \mysou, with an estimated 
5$\sigma$ upper limit on the source flux of 4\,mCrab in the 3--10\,keV energy band and of 2\,mCrab in the 10--20\,keV energy band.\\

At the time of writing, the nature of \mysou~is still to be unveiled. In this paper we present the results of our \chandra~observation as well as long term \integral~and \swift~observations, to place our \chandra~observation in the source emission context, as well as to obtain a broad-band source coverage. We report also on archival and new NIR observations of the source taken during the outburst.

\section{Observations and data analysis}

\subsection{\chandra~data}\label{sec:chandra}

We observed \mysou~for 20\,ks with \chandra~on 2014 November 3, between UT 00:05 and 06:17 (MJD 56964.0-56964.26, Observation ID 15744) with the High Energy Transmission Grating Spectrometer, HETGS \citep{canizares00}.

Throughout this work we
shall consider data from the \zth\,spectral order for source position extraction, and from the $\pm$\fst\ orders for spectral extraction. 
Higher spectral orders have very low count rates and thus shall be ignored, while the \zth\,order spectrum will not be considered in the spectral analysis as it severely suffers from pileup.  The data were analyzed in a standard manner, using the CIAO version 4.6 software package and \chandra CALDB version 4.6.3. 

To increase the signal-to-noise ratio, we have merged the $\pm$\fst\, orders in a single first order Medium Energy Grating (MEG) spectrum and a single first order High Energy Grating (HEG)  spectrum. Starting from 0.5\,keV and 0.7\,keV in \meg~and \heg, respectively, the data were grouped to have 7$\sigma$ bins (for the investigation of discrete features) and 14$\sigma$ (for the joint \chandra/\swift~spectra). Spectra were fitted using {\tt{XSPEC}} version 12.7.0. Only the results from the latter grouping (consistent with the former) are shown in the paper. \\

Our \chandra~observation revealed another bright source located slightly further than 18$^{\prime}$ from our target source. Basic results on the source are also given, in Section~\ref{sec:chandranew}.

\subsection{\swift~data}\label{sec:swift}

The \swift~satellite pointed \mysou~five times between 2014 October 11 and 2014 November 2, the latter being one day prior to our 
\chandra~observation. The log  of the observations is reported in Table~\ref{tab:swiftdata}.

\begin{table}[htbp]
\caption{Journal of the \swift/XRT observations of \mysou.}
\begin{tabular}{cccc}
ObsId & Date Start & Exposure & XRT mode \\
  (\#)         &      (UTC)            &    (s)         &                   \\
00033470002 & 2014-10-11 00:10:17 & 4640 & WT\\
00033470003 & 2014-10-13 06:56:17 & 1594 & WT\\
00033470004 & 2014-10-13 04:48:45 & 4365 & PC$^\star$\\
00033470006 & 2014-10-23 08:06:21 & 1673 & WT\\
00033470007 & 2014-11-02 07:52:01 & 1015 & PC\\
\end{tabular}
\begin{list}{}{}
\item[$^\star$]Spectrum strongly piled-up and not used.
\end{list}
\label{tab:swiftdata}
\end{table}

The \swift/XRT spectra were obtained thanks to
the on-line tool provided by the \swift~UK center\footnote{http://www.swift.ac.uk/user\_objects/}.
The complete procedure for products extraction is described in \citet{evans09}. Note that, as 
recommended for absorbed sources, we extracted spectra from grade 0 only for the window timing 
data. We also cross-checked the results by reducing a couple of the observations from the raw data 
following standard procedures using the \heasoft~software suite \citep[through \xselect, e.g.][]{Rodriguez10,Rodriguez11}. As the products
showed no significant deviations, we used those obtained 
from the on-line tool for the spectral fits obtained here. 
The data were grouped so as to have a minimum of 25 counts/bin and then fitted 
in {\tt{XSPEC}} between 0.6\,keV and 8/10\,keV, depending on the quality of the data.

The \swift/BAT survey data, retrieved from the HEASARC public 
archive\footnote{http://swift.gsfc.nasa.gov/archive/}, cover the period from MJD 56778.2 to 57032.9 (2014 May 1 -- 2015 January 10). 
Due to Solar constraints, no observatory could look at the source after 2014 November and \mysou~was not detected in the available \swift/BAT data after the solar constraint, up to 2015 January 10.

The \swift/BAT data 
were processed using {\tt BAT\_IMAGER} software \citep{segreto2010}.
This ad-hoc code, dedicated to the processing of coded mask instrument data,
computes all-sky maps  and, for each detected source, produces standard products such as light curves and spectra. We note that the code
takes into account the cross-contamination between sources in the field of view. This is essential in crowded fields such as the one of \mysou~that has two nearby sources, 1A~1742--294, 13$^{\prime}$ away, and AX~J1745.6$-$2901,  18$^{\prime}$ away \citep[see also][for another application of the decontamination process]{cusumano15}. The \swift/BAT spectra were analyzed in {\tt{XSPEC}} between 15\,keV and 150\,keV.

\subsection{\integral~data}\label{sec:integral}

\mysou~has been in the \integral~field of view during the observations of the {\sl Galactic Centre} (ID 1120027)  and of the  {\sl Galactic Bulge region\footnote{http://integral.esac.esa.int/BULGE/}} (ID 1120001). The third set of observations covering the source, ID 1020021, consists of proprietary data and is not included in this work.
A complete study of these \integral~data is out of the scope of this paper, however to have a feeling of the broad-band 
long-term behavior of the source,
we have analyzed the IBIS/ISGRI and JEM--X\footnote{Results are given for JEM--X1 and JEM--X2 combined.} data \citep[][respectively]{lebrun03,lund03} starting from revolution 1446
(2014 August 18, 01:17:28 UT, MJD 56887.05, about forty days prior to the reported discovery) up to revolution 1470 (2014 October 28, 15:42:00 UT, MJD 56958.65), the latest dates compatible with solar constraints. The first \integral~observation of the Galactic center occurred again on 2015 February 16--17, during which \mysou~was not detected \citep{boissay15}.

A standard analysis using version 10.1 of the Off-line Scientific Analysis (\osa) software
was performed on the pointings where the source was simultaneously in the field of view of IBIS/ISGRI and JEM--X.

\subsection{Near-infrared data}\label{sec:NIR}
With our \chandra~position of \mysou~at hand, we searched for candidate counterparts in several NIR surveys as well as obtained new optical/NIR observations of the source field during the outburst.

The position of \mysou~had been observed on 2010 August 15 by the VISTA Variables in the V\'ia L\'actea Survey \citep[VVV,][]{minniti10}, on 2006 July 18 during the UKIRT Infrared Deep Sky Survey \citep[UKIDSS\footnote{The UKIDSS project is defined in \citet{lawrence12}. UKIDSS uses the UKIRT Wide Field Camera \citep[WFCAM;][]{casali07}. The photometric system is described in \citet{hewett06}, and the calibration is described in \citet{hodgkin09}. The pipeline processing and science archive are described in \citet{hambly08}.  We  
have used data from the 10th data release, which is described in detail in \citet{lawrence12}.},][]{lawrence07}, and also on 1998 July 2 as part of the Two Micron All-Sky Survey \citep[2MASS,][]{skrutskie06}.

Furthermore, we observed \mysou~during the outburst with the 7-channel imager Gamma-Ray burst Optical Near-infrared Detector \citep[GROND,][]{greiner08} at the MPG 2.2\,m
telescope at ESO La Silla Observatory at UT 00:31 on 2014 October 6 (MJD 56936.0).  
GROND observes in the four optical and three near-IR bands simultaneously. Because of the severe Galactic foreground
reddening (A$_{V}\sim$44, A${_J}\sim$12), we consider here only the $J$, $H$, and $K_s$ bands. The total integration times in each band was $10.7$\,min and the average seeing was $1.3^{\prime\prime}$.
 The GROND data were reduced and analyzed with the standard tools and methods described in \citet{kruhler08}. 
Here, the $J,H$ and $K_s$ photometry was measured from 1.3$^{\prime \prime}$ apertures and calibrated relative to point-like field sources from the UKIDSS DR10 \citep{lawrence12}.

\section{Results}\label{sec:results}
\begin{figure}

\includegraphics[width=1.0\linewidth]{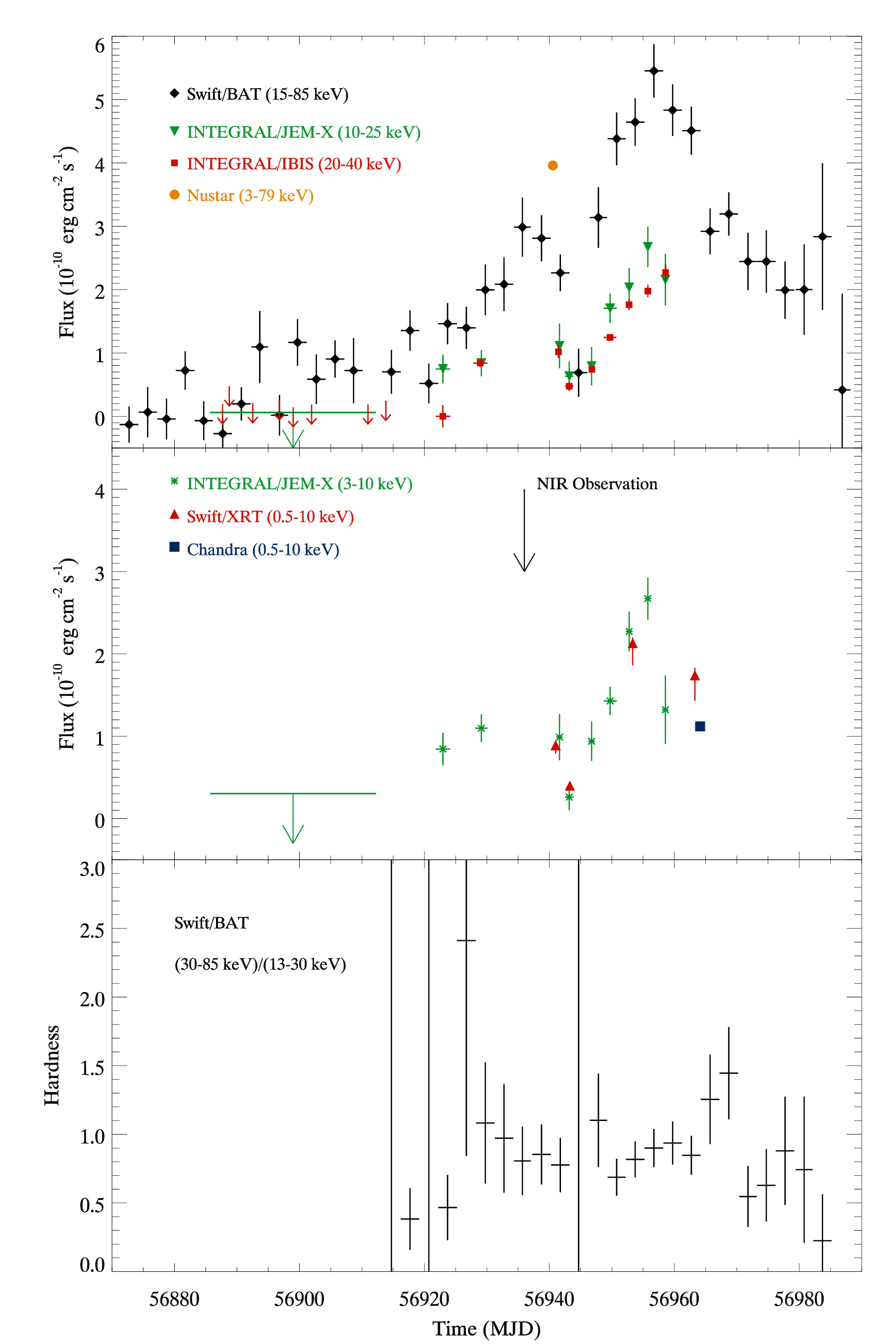}

\caption{The 2014 outburst of \mysou~as seen by different high energy missions. The date of our GROND NIR observation is marked as a black vertical arrow in the middle panel. See Section~\ref{sec:results} for details. }
\label{fig:allLCR}
\end{figure}

A summary of all the data discussed in this work is shown in Figure~\ref{fig:allLCR} where the overall 2014 outburst of \mysou~can be seen.

The upper panel depicts the \emph{hard X--ray} behavior of \mysou~with results from 
\integral/JEM--X (10--25\,keV flux, green down triangles), \integral/IBIS (20--40\,keV flux, red boxes)
and \swift/BAT (15--85\,keV flux, black filled diamonds, 3\,day bins). The \textit{NuStar} unabsorbed 3--79\,keV flux 
\citep[yellow circle,][]{tendulkar14} is also shown here.

The middle panel shows the \emph{soft X--ray} behavior with results from 
\integral/JEM--X (absorbed 3--10\,keV flux, green asterisks), \swift/XRT 
(absorbed 0.5--10\,keV flux, red triangles), \chandra/HETGS (absorbed 0.5--10\,keV flux, blue box). 
The time of our NIR GROND observation is marked by the black vertical arrow.
As it can be seen, it occurred during the first peak of the outburst of \mysou, as traced by \swift/BAT (upper panel).

The lower panel is the hardness ratio evolution computed with \swift/BAT in 30--85\,keV versus 15-30\,keV.

\subsection{\chandra~results}

\subsubsection{\chandra position, variability and spectra of \mysou}\label{sec:chandrares}

\begin{figure}
\includegraphics[width=0.77\linewidth,angle=270]{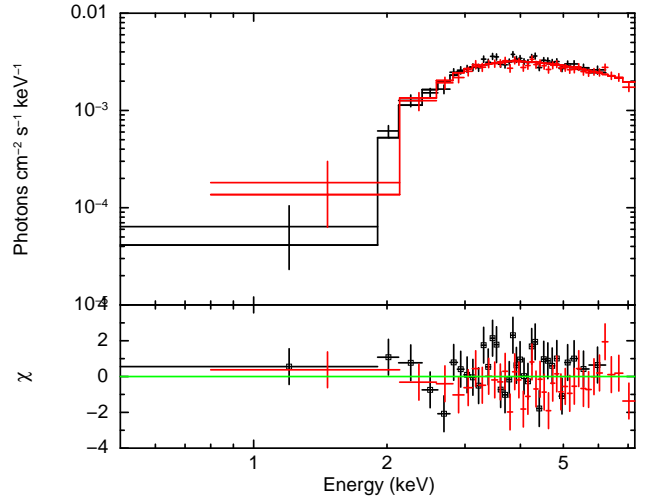}
\caption{Merged $\pm$\fst\,order MEG (black, box symbol in the residuals) and HEG (red) spectra with best fit {\tt tbabs*po} model. See Table~\ref{tab:allfits} and  Section~\ref{sec:chandrares}.}
\label{fig:chandraspe}
\end{figure}

Given the brightness of the source, the \zth\,order image is piled-up, hence the resulting shape of the point spread function is distorted and no longer exactly point-like. This renders more difficult to locate the precise centroid position of the source. For this reason, the source location was determined by intersecting the readout streak with the grating arms. This was accomplished with the  \texttt{findzo} algorithm, which is  used for determining the \zth\, order position when a readout streak is detected (and hence pileup is affecting the \zth\,order image), as in our case.  

The X-ray position obtained is  $\alpha_\mathrm{J2000}$=17$^\mathrm{h}$ 45$^\mathrm{m}$ 27$^\mathrm{s}$.69, 
\mbox{$\delta_\mathrm{J2000}$= $-$29$^{\circ}$ 19$^{\prime}$ 53$^{\prime \prime}$}.8, as reported in \citet{paizis15}.
Since the statistical error is smaller than the 
absolute position accuracy of \chandra, 0$^{\prime \prime}$.6 at 90\% 
uncertainty\footnote{http://cxc.harvard.edu/cal/ASPECT/celmon/},  we attribute to the position 
found a 90\% uncertainty  of 0$^{\prime \prime}$.6. 

The \chandra~light-curve of \mysou, similarly to what was reported by  \citet{chenevez14} and \citet{tendulkar14}, showed no evidence for bursts or pulsations on the time scales accessible to our \chandra~observation, i.e. $\approx$4--10000\,s (twice  a  \chandra~time bin to half the observation).

The \chandra~spectrum of \mysou~can be well fit 
by an absorbed ({\tt tbabs}) power-law  (${\rm N_H}\sim$12$\times 10^{22}\,{\rm cm}^{-2}$, and photon index 
$\Gamma\sim$1.6) with
an average absorbed 2--8\,keV flux of about 1$\mathrm{\times 10^{-10}~erg~cm^{-2}~s^{-1}}$.
Figure~\ref{fig:chandraspe} shows the best fit we obtained with the absorbed power-law model, while Table~\ref{tab:allfits} shows the obtained parameters.

In the fit, and throughout the paper, we have used an improved model for the absorption of X-rays in the ISM by 
\citet{wilms00}\footnote{In XSPEC terminology: {\tt tbabs} with {\tt xsect vern} and {\tt abund wilm}.}. Such a model results in higher column densities with respect to, e.g., the {\tt wabs} model by \citet{morrison83}.
For comparison, using the {\tt wabs} model, the following is obtained: ${\rm N_H}$=(8.2$\pm$0.7)$\times 10^{22}\,{\rm cm}^{-2}$, and photon index $\Gamma$ =1.6$\pm$0.17.

%%%%%%%%%%%%%%%%%%%%%%%%%%%%%%%%%%%%%%%%%%%%%%%%%%%%%%%%%
\begin{deluxetable*}{lcccccc}
\setlength{\tabcolsep}{0.03in} 
\tabletypesize{\footnotesize}    
\tablewidth{0pt} 
\tablecaption{Fit to \mysou~spectra: {\tt tbabs}*{\tt po}.\label{tab:allfits}}
\tablehead{
  \colhead{Spectra} & \colhead{$N_\mathrm{H}$\tablenotemark{(a)}}
   & \colhead{$\Gamma$} 
   & \colhead{Average U\_Flux} 
& \colhead{Average Flux}
& \colhead{Average Luminosity}
   & \colhead{Red $\chi^2/$Dof}
          \\                               
 &  
 ($10^\mathrm{22}~\mathrm{cm^\mathrm{-2}}$) 
   & 
   & ($\mathrm{10^{-10}~erg~cm^{-2}~s^{-1}}$)
  & ($\mathrm{10^{-10}~erg~cm^{-2}~s^{-1}}$)
&   (10$^{36}$ $\mathrm{erg~s^{-1}}$)     }
\startdata 
\\
 \chandra/HETGS 
 & \errtwo{12.1}{0.8}{1.1}     		 % N_H
 & 1.6$\pm$0.2     % Powerlaw Slope
 & 2.0\tablenotemark{(b)}     % Unabsorbed flux
 & 1.1\tablenotemark{(c)}  % Absorbed flux
 & 0.9\tablenotemark{(d)}    % Absorbed L
 & 1.05/66               \\      % Chi^2/DoF
\\
\swift/(XRT$+$BAT) 
& \errtwo{10.5}{1.2}{1.1}  	 % N_H
& 1.8$\pm$0.1 % Powerlaw Slope
& 9.7\tablenotemark{(e)}   % Unabsorbed flux
&  7.1\tablenotemark{(f)} % Absorbed flux
& 5.4\tablenotemark{(g)} % Absorbed L
&  0.87/66   \\
\\
\chandra/HETGS$+$\swift/BAT 
&\errtwo{11.9}{1.1}{1.0}    % N_H
&  1.6$\pm$0.2   % Powerlaw Slope
&  11.7\tablenotemark{(e)} % Unabsorbed flux
& 9.7\tablenotemark{(f)}  % Absorbed flux
 & 7.4\tablenotemark{(g)}   % Absorbed L
& 1.03/69   \\

  \tablecomments{Error bars are 90\% confidence level for one 
     parameter. \chandra~spectra are shown in Figure~\ref{fig:chandraspe} and discussed in Section~\ref{sec:chandrares}.
\swift/(XRT$+$BAT) spectra are shown in Figure~\ref{fig:jointspe}, upper panel, and discussed in Section~\ref{sec:swiftfits} while 
\chandra$-$\swift/BAT spectra are shown in Figure~\ref{fig:jointspe}, lower panel, and discussed in Section~\ref{sec:swiftchandrares}.}
\tablenotetext{(a)}{In the fit we have used an improved model for the absorption of X-rays in the interstellar medium by \citet{wilms00}.} 
\tablenotetext{(b)}{Unabsorbed 2--8\,keV flux.} 
\tablenotetext{(c)}{Absorbed 2--8\,keV flux.} 
\tablenotetext{(d)}{Absorbed 2--8\,keV luminosity, assuming a distance of 8\,kpc.} 
\tablenotetext{(e)}{Unabsorbed 0.5--100\,keV flux.} 
\tablenotetext{(f)}{Absorbed 0.5--100\,keV flux.} 
\tablenotetext{(g)}{Absorbed 0.5--100\,keV luminosity, assuming a distance of 8\,kpc.} 

\end{deluxetable*}
%%%%%%%%%%%%%%%%%%%%%%%%%%%%%%%%%%%%%%%%%%%%%%%%%%%%

\subsubsection{The second source in the \chandra~field of view: AX~J1745.6$-$2901 }\label{sec:chandranew}
Our \chandra~observation revealed another bright source 
located $\sim$18$^{\prime}$ from our target source\footnote{We note that this source does not contaminate the results we obtain for
\mysou~in hard X--rays, neither with \integral/IBIS-JEM--X nor with \swift/BAT. Indeed, for \integral, the source is further than the instrncic instrumental angular resolution (12$^{\prime}$ for IBIS and 3$^{\prime}$ for JEM--X), while for BAT source  cross-contamination is taken care of by the code used (see Section~2.2).}.  
At this off-axis angle, the \chandra~point spread function (PSF) becomes
very extended and obtains an elliptical shape.  The image of this
source appears as an ellipse with semi-major/semi-minor axes of
approximately 1.1$^{\prime}$/0.7$^{\prime}$, respectively, with shadows of the
mirror support structures clearly visible in the PSF.  The
intersections of these shadow structures occur within 2$^{\prime \prime}$ of
$\alpha_\mathrm{J2000}$=17$^\mathrm{h}$ 45$^\mathrm{m}$ 35$^\mathrm{s}$.44, 
\mbox{$\delta_\mathrm{J2000}$= $-$29$^{\circ}$ 01$^{\prime}$ 33$^{\prime \prime}$}.6 (J2000), which is the location of a
known transient, the bursting NS AX~J1745.6$-$2901  \citep[see][and references therein]{degenaar2009, muno2004}.

We extracted the \zth\,order spectrum for this source using the CIAO
\texttt{specextract} tool (taking the background from a nearby 0.7$^{\prime}$
circular region).  Binning the spectrum to a signal-to-noise per
channel of 8 and noticing the 2--8\,keV region, the spectrum is well
fit ($\chi^2$=224.6 for 222 degrees of freedom) by an absorbed, $N_H = (34\pm2) \times 10^{22}\,{\rm
  cm^{-2}}$  power-law
($\Gamma=1.96\pm0.14$).  Error bars are 90\% confidence for
one interesting parameter.  This $N_H$ value is 50\% higher than the
value reported by \citet{degenaar2009}; however, they do not
specify the absorption model used.  

The fitted, unabsorbed 2--10\,keV luminosity  is $6.1\times10^{36}\,{\rm
  erg\,cm^{-2}\,s^{-1}}$, assuming isotropic emission at a distance
of 8\,kpc.  This would place this outburst of AX~J1745.6$-$2901 at
the same level as the brightest of the four historical outbursts
discussed by \citet{degenaar2009}, with the other outbursts
being approximately 6--30 times fainter.

We searched the source light-curve for evidence of variability (there
is an 8.4\,hr period known from eclipses), including any evidence of
type-I bursts, but no statistically significant variability was found
on any time scale accessible to this \chandra~observation ($\approx$4--10000\,s).

\begin{figure}
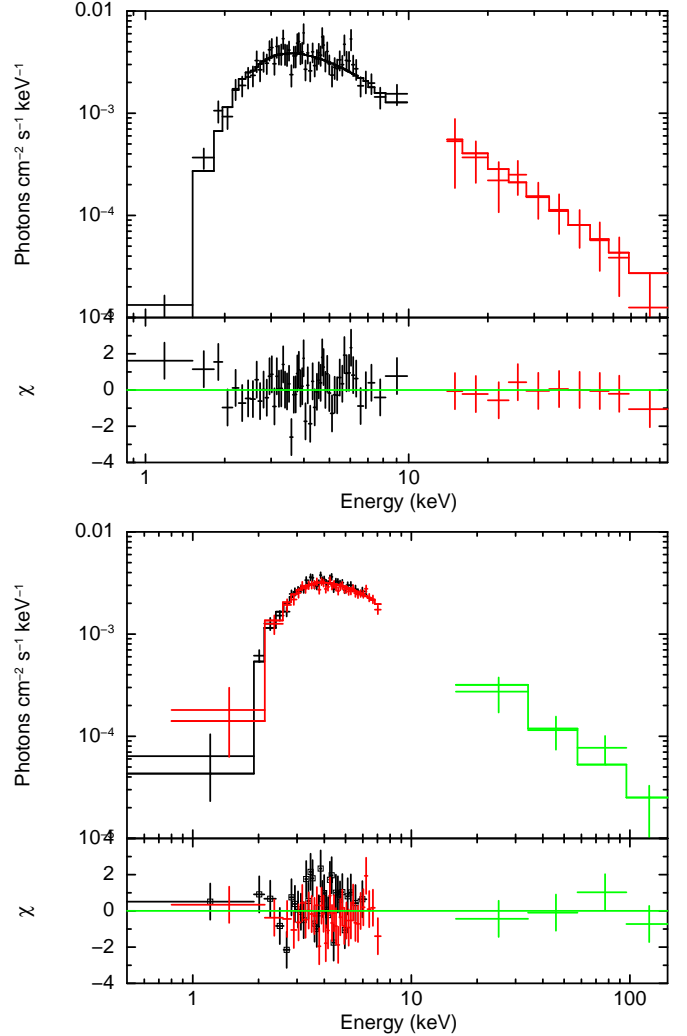

\includegraphics[width=0.8\linewidth,angle=270]{SWIFT_po.ps}
\includegraphics[width=0.8\linewidth,angle=270]{simult_po.ps}
\caption{{\it Upper panel}: \swift/XRT$+$BAT observation of \mysou~(November 2, ObsID 00033470007) and best fit model {\tt tbabs*po}. See Table~\ref{tab:allfits} and Section~\ref{sec:swiftfits}.
{\it Lower panel}: joint \chandra-\swift/BAT~spectra of \mysou~(November 3). Merged $\pm$\fst\,order MEG (black), HEG (red), \swift/BAT (green) spectra with best fit {\tt tbabs*po} model. See Table~\ref{tab:allfits} and Section~\ref{sec:swiftchandrares}.}
\label{fig:jointspe}
\end{figure}

\subsection{\swift~results on \mysou}\label{sec:swiftres}
\subsubsection{\swift/XRT results}
Excluding the severely piled up observation 00033470004 (see Table~\ref{tab:swiftdata}), we have extracted \swift/XRT spectra for the remaining four observations. All spectra could be well fit by an absorbed power-law. The obtained fluxes are shown in Figure~\ref{fig:allLCR}, middle panel. During three of these four observations (first, third and fourth, chronologically), the source did not show any relevant spectral evolution although the source flux doubled, with spectral slope remaining around $\Gamma$=1.5$\pm$0.3 and absorption within the interval ${\rm N_H}\sim$(7.5-10.7)$\times 10^{22}\,{\rm cm}^{-2}$. The second observation, 00033470003 occurred on 2014 October 13 (MJD=56943.28), resulted in a very poorly constrained fit ($\Gamma$=\errtwo{2.1}{1.4}{1.0} and ${\rm N_H}$=\errtwo{8.5}{6.5}{3.7}$\times 10^{22}\,{\rm cm}^{-2}$) since the source experiences a flux drop, as also detected by \integral~and \swift/BAT (Figure~\ref{fig:allLCR}).\\
\\

\subsubsection{\swift/BAT results}\label{sec:swiftbatresadded}
Due to its extremely wide field of view, \swift/BAT has nicely monitored \mysou. The all-outburst \swift/BAT light-curve (3 day bins) can be seen in Figure~\ref{fig:allLCR}, upper panel, while the hardness ratio in the bands 30-85\,keV versus 15--30\,keV  is shown in Figure~\ref{fig:allLCR}, lower panel. 

A complete study of all the \swift/BAT data is beyond the scope of the paper, but we have extracted \swift/BAT spectra for three time intervals: the first one corresponding to \swift/XRT ObsID 00033470007 (see Table~\ref{tab:swiftdata}, MJD 56963.3, November 2, \swift/BAT exposure of about 87\,ks) about a day prior to our \chandra~data, the second  \emph{simultaneous} to our \chandra~observation (MJD 56964.0, November 3, $\sim$17\,ks), and the third covering the outburst peak of \mysou, from October 20 to November 4 (MJD 56950--56965, roughly the highest five bins in Figure~\ref{fig:allLCR}, 172\,ks). 

The three \swift/BAT spectra could be well fit by a simple power-law with $\Gamma$=1.9$\pm$0.5 (November 2), $\Gamma$=1.3$\pm$0.8 (November 3) and  $\Gamma$=1.9$\pm$0.1 (peak). Though the second result ($\Gamma$=1.3$\pm$0.8) appears to suggest a hardening of \mysou, the result is not statistically significant and the slopes overlap.

In the next Section we use the  \swift/BAT spectra to investigate the broad-band behavior of \mysou.

\subsection{Broad-band results of \mysou}
\subsubsection{\swift/(XRT$+$BAT) observation\\ (November 2)}\label{sec:swiftfits}

The simultaneous \swift/(XRT$+$BAT) spectrum of \mysou~as obtained from November 2 (see Table~\ref{tab:swiftdata}) can be well fit  by an absorbed power-law  (${\rm N_H}\sim$10.5$\times 10^{22}\,{\rm cm}^{-2}$ and photon index $\Gamma\sim$1.8) with
an average absorbed 0.5--100\,keV flux of about 7.1$\mathrm{\times 10^{-10}~erg~cm^{-2}~s^{-1}}$.
Figure~\ref{fig:jointspe}, upper panel, shows the best fit we obtained with the absorbed power-law model, while Table~\ref{tab:allfits} shows the obtained parameters. There is clearly no need for a power-law cut-off energy in the data.
\\
\\
\\

\subsubsection{\chandra-\swift/BAT observation \\(November 3 and peak)}\label{sec:swiftchandrares}

As visible in Figure~\ref{fig:allLCR}, our \chandra~observation is simultaneous to a \swift/BAT coverage. Hence to obtain broad-band information on the source, we performed a joint \chandra-\swift/BAT~spectral fitting of the simultaneous data. 
 
The data can be well fit by an absorbed power-law  (${\rm N_H}\sim$11.9$\times 10^{22}\,{\rm cm}^{-2}$, and photon index $\Gamma\sim$1.6) with
an average absorbed 0.5--100\,keV flux of about 9.7$\mathrm{\times 10^{-10}~erg~cm^{-2}~s^{-1}}$.
Figure~\ref{fig:jointspe}, lower panel, shows the best fit we obtained with the absorbed power-law model, while Table~\ref{tab:allfits} shows the obtained parameters.

Similarly to the previous case, there is no trace of a high energy cut-off in the data.

From Figure~\ref{fig:allLCR} it is possible to see that our \chandra~observation occurred at a similar \swift/BAT (15--85\,keV) flux as the \nustar~one. Hence to compare the results, we fitted our broad-band \chandra-\swift/BAT spectra with the same model used by \citet{tendulkar14}, i.e. a cut-off power-law with absorbing model  ({\tt tbabs}, Tendulkar, private communication) and abundances by \citet[][]{anders89}, instead of the ones used up to now by \citet{wilms00}. 
We obtained: ${\rm N_H} =7.7 \pm0.7 \times 10^{22}\,{\rm cm}^{-2}$, $\Gamma$ =\errtwo{1.57}{0.07}{0.18} and cut-off energy $E_{cut}>$80\,keV. Compared to the results by \citet{tendulkar14} (${\rm N_H} =3.3 \pm0.6 \times 10^{22}\,{\rm cm}^{-2}$, $\Gamma = 1.46\pm0.06$ and $E_{cut}>$100\,keV), we see that while the power-law slope and cut-off energy are comparable, the absorbing column density is about a factor of two different.

As a final check, we fit the \chandra~data with the peak \swift/BAT spectrum (with the \emph{peak} defined in Section~\ref{sec:swiftbatresadded} i.e. from October 20 to November 4, roughly the highest five bins in Figure~\ref{fig:allLCR}), in an attempt to improve the fit quality with the better \swift/BAT statistics, for there is an overlap, though the data are not exactly simultaneous. Similarly to the previous cases, there is no significant detection of a cut-off in the spectrum with energy $E_{cut}>$80\,keV.

\subsection{\integral}\label{sec:integralres}
During the discovery outburst, the source was never detected in a single \integral~pointing ($\sim$2\,ks) and several pointings needed to be stacked to increase the sensitivity with longer exposures. Therefore,  we have built mosaic images with the IBIS/ISGRI and JEM--X1$+$2 instruments, to obtain average flux
measurements for each \integral~revolution. In Figure~\ref{fig:allLCR}, we show, in the upper panel, the source flux variations between 20--40\,keV (IBIS/ISGRI) and 10--25\,keV (JEM--X), while the 3--10\,keV (JEM--X) flux variations are shown in the middle panel.
In the case of JEM--X, upper limits in Figure~\ref{fig:allLCR} are from the total combined JEM--X mosaic with the deepest exposure available prior to the onset of the outburst and are at 5$\sigma$, while for IBIS/ISGRI the upper limits are given per revolution and are at 3$\sigma$. 

Inspection of our all-public IBIS/ISGRI archive \citep[October 2002 -- March 2014;][17--50\,keV]{paizis2013} for previous unnoticed outbursts from \mysou~shows that during the 35.2\,Ms (good IBIS/ISGRI time) in which \mysou~was in the IBIS/ISGRI field of view (source position within 15$^{\circ}$ from the pointing coordinates), the source has never been detected at a single pointing level ($\sim$2\,ks), implying that \mysou~was (at best) at the detection limit in IBIS/ISGRI, corresponding to about 20\,mCrab in a 2\,ks pointing \citep{krivonos2010}.

\begin{figure}

\includegraphics[width=1.0\linewidth]{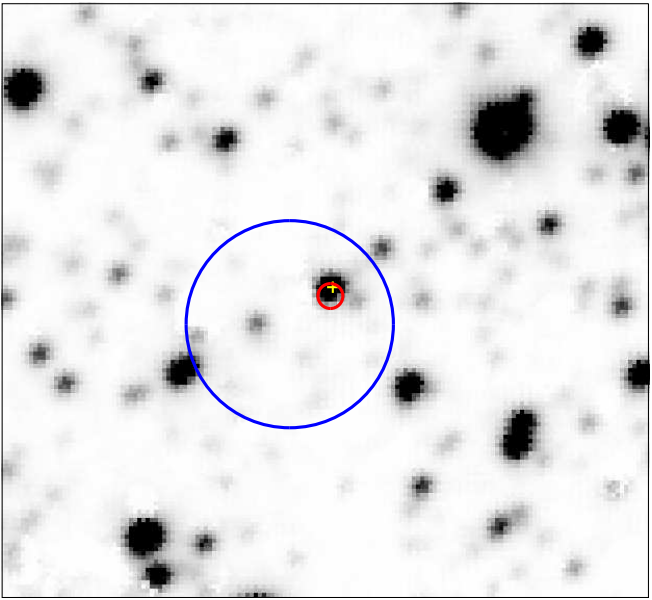}
\caption{Archival UKIDSS $K_{s}$-image of the field around \mysou. The \swift/XRT 90\% error circle \citep[5$^{\prime \prime}$ in blue,][]{chenevez14} and 
\chandra~90\% error circle (0.6$^{\prime \prime}$ in red, this work) are shown. 2MASS~J17452768$-$2919534 is indicated with a yellow cross. North is top, East is left. See Section~\ref{sec:nirres}.}
\label{fig:Kband}
\end{figure}

\subsection{Near-infrared}\label{sec:nirres}
Figure~\ref{fig:Kband} shows the archival UKIDSS $K_{s}$-image of the field around \mysou. The \swift/XRT 90\% error circle \citep[5$^{\prime \prime}$ in blue,][]{chenevez14} and \chandra~90\% error circle (0.6$^{\prime \prime}$ in red, this work) are shown.
While the \swift/XRT position error includes more than one source in the crowded field towards the  Galactic center, our \chandra~position coincides within the errors with a 2MASS object, 2MASS~J17452768$-$2919534 (indicated with a yellow cross in the map) which makes it the most likely NIR counterpart.

Due to the large Galactic foreground reddening, no optical counterpart is
detected in the GROND $g^\prime, r^\prime, i^\prime, z^\prime$ bands. 
However the likely 2MASS counterpart is detected in the infrared $J, H$ and $K_s$ bands. 

A comparison of the GROND, UKIDSS, 2MASS and VVV\footnote{We took here default 2$^{\prime \prime}$ aperture photometry of the VVV catalogue (so-called \texttt{jAperMag3}, \texttt{hAperMag3} and \texttt{ksAperMag3}), and we also added the uncertainty in the photometric zero-points of 0.01\,mag, in quadrature to the catalogued statistical uncertainties.} photometry (all magnitudes in the Vega System) is presented in Table~\ref{tab:GROND-UKIDSS-2MASS-VISTA}.
\begin{table}
\caption{GROND, 2MASS, UKIDSS and VVV photometry}
\label{tab:GROND-UKIDSS-2MASS-VISTA}
\begin{tabular}{ccccc}
\hline
Band       & GROND$^a$      & 2MASS$^b$       & UKIDSS$^b$           & VVV$^b$ \\
\hline
$J^c$      & $16.07\pm0.08$ & $>16.227$         & $16.587\pm0.015$  & $16.46\pm0.02$ \\
$H$        & $13.10\pm0.07$ & $13.038\pm0.065$ & $13.150\pm0.003$  & $13.12\pm0.01$ \\
$K_{s}$    & $11.37\pm0.06$ & $11.365\pm0.024$ & $11.334\pm0.002$  & $11.37\pm0.01$\\
\hline
\end{tabular}
\raggedright$^a$: NIR during source outburst.\\
\raggedright$^b$: Archival NIR catalogues.\\
\end{table}
The candidate counterpart is constant within the error bars in the $K_{s}$-band and also constant between GROND, UKIDSS and VVV in the $H$-band. 
The $J$ band, however, shows a hint of variability with respect to the UKIDSS and VVV survey.
 Hence, we conclude that 2MASS~J17452768$-$2919534 underwent  a slight brightening in the $J$-band during the X--ray outburst, while it remained constant in the  $H$ and $K_s$ bands.

\section{Discussion}\label{sec:discussion}
After the discovery, \mysou~has undergone an outburst of about 60 days (MJD 56923.7-56983.7, as suggested by the last non-detection in \swift/BAT, see Figure~\ref{fig:allLCR}).
Due to Solar constraints, however, no observatory could look at the source after 2014 November, so a longer outburst cannot be excluded. 

The X--ray light-curve of \mysou~(Figure~\ref{fig:allLCR}) shows a double peak with a clear flux decrease seen
in all X--ray bands around MJD 56945 (duration $\sim$4--5\,d). The
obtained spectra during such a dip are too dim to verify a significant variability of the spectral model parameters (see Section~\ref{sec:swiftres} and
Fig~\ref{fig:allLCR} lower panel), but the fact that this flux decrement is present in all
the X--ray data may suggest that it does not correspond to an important spectral
state change, rather to a decrease in the  overall flux. Such a decrease
may be due to an intrinsic change in mass transfer from the companion in
an eccentric orbit \citep[see, e.g. the few, $\sim$2--3, day dip in the
X--ray light-curve of Cir~X--1,][]{shirey96, murdin80}, or due to a change
of the absorbing medium \citep[dips from intervening matter or eclipses by
the donor or by a tilted and/or warped accretion disk, see e.g.][and
references therein]{clarkson03}.

Broad band spectra of the source are compatible with an absorbed power-law ($\Gamma\sim$1.6-1.8 and ${\rm N_H}\sim$10--12$\times 10^{22}\,{\rm cm}^{-2}$), with no trace of a cut-off in the data. Unlike \nustar~\citep{tendulkar14}, we do not see evidence of a broad iron line in our soft X--ray spectra, which is not a surprise considering the  significantly larger effective collecting area in the iron domain of \textit{NuSTAR} with respect to \chandra~and \swift/XRT. 
Using the same model by \citet{tendulkar14} to fit our broad-band \chandra-\swift/BAT spectra, we obtained a significantly different absorbing column density, ${\rm N_H} =(7.7\pm0.7) \times 10^{22}\,{\rm cm}^{-2}$ versus their ${\rm N_H} =(3.3 \pm0.6) \times 10^{22}\,{\rm cm}^{-2}$. This could imply a real variability in ${\rm N_H}$ (hence a part of it is local to the system), but we note that  the discrepancy could be due to the lack of data below 3\,keV and/or to the modeling of the Fe line in the \nustar~data.

The X--ray characteristics of \mysou~are typical of a Low Mass X--ray Binary (LMXB) rather than of a High Mass X--ray binary (HMXB). Indeed, most HMXBs, 
apart from the exceptionally few BH HMXBs such as Cyg~X--1, have spectra with a very clear cut-off below 40\,keV \citep{coburn02}, while \mysou~has a clear non attenuated power-law up to 100\,keV, as seen in the broad band spectra presented in the present work. Furthermore, the spectral slope of \mysou~is typical of an LMXB in the low-hard state,  different from
the flatter slope  generally seen in HMXBs ($\Gamma\sim$1).

Regarding the nature of the compact object in the binary system, we note that up to now no pulsations or type-I X-ray bursts, that would point to the presence of a NS in the system, have been detected \citep[this work,][]{chenevez14,chenevez14b,tendulkar14}.  Furthermore, the spectral analysis and long term light-curve investigated in this work do not seem to strongly favor the BH option, with respect to the NS. Indeed, the ultra-compact binary and X--ray burster (hence an LMXB with NS) 4U~1850--087 has shown a similar spectrum to \mysou: a non attenuated power-law up to about 100\,keV,  with best-fit photon index of $\Gamma$=1.9$\pm$0.1 and 2--100\,keV luminosity of  $\sim$1.5$\mathrm{\times 10^{36}~erg~s^{-1}}$ at 6.8\,kpc  \citep{sidoli06}. This  source, observed with \textit{BeppoSAX} and \integral, spent most of the time in this low luminosity hard state.
Its simultaneous \xmm/EPIC-\integral/IBIS spectrum required a soft disk emission besides the power-law ($kT_s$=0.8$\pm$0.1), but the source had a very low absorbing column density ($\rm N_H$=0.4$\times 10^{22}\,{\rm cm}^{-2}$), whereas in our case $\rm N_H$, be it local to the system or Galactic, is high and the soft disc component, if any, is most likely hidden and undetected.

The peak flux reached by \mysou~during the outburst as seen by \swift/BAT in 15--85\,keV is
(5.5$\pm$0.4)$\mathrm{\times 10^{-10}~erg~cm^{-2}~s^{-1}}$ (Figure~\ref{fig:allLCR}). Placing the source at 8\,kpc and assuming the same spectrum as in  Table~\ref{tab:allfits}, we obtain a corresponding 0.5--100\,keV peak luminosity of 9$\mathrm{\times 10^{36}~erg~s^{-1}}$  (3.5$\mathrm{\times 10^{36}~erg~s^{-1}}$ at 5\,kpc) that is in the range of X--ray binary outbursts (both LMXB and HMXB). In this respect, the peak flux does not
allow us to constrain the nature of the source. On the other hand, broad asymmetric iron lines as the one detected by \citet{tendulkar14}  are in general typical of accretion disks, hence mainly found in LMXBs \citep[e.g.][]{ng10}, whereas in the case of HMXBs, the observed lines have normally a narrow profile (e.g. Vela~X--1,
GX~301--2, 4U~1700--37), usually interpreted as fluorescence
of iron in a wind or circumstellar matter  \citep[][]{rodriguez06, gimenez15}.

The large value of the absorption we obtain (${\rm N_H}\sim$10--12$\times 10^{22}\,{\rm cm}^{-2}$ or ${\rm N_H}$=8.2$\pm$0.7$\times 10^{22}\,{\rm cm}^{-2}$ using \texttt{wabs}) is well in excess with respect to the average Galactic value in the source direction, $\sim$1.2$\times10^{22}$\,cm$^{-2}$ \citep{dickey90}. This could imply that there is an additional contribution from within the system and/or that we are seeing the system at high inclination. Nevertheless,
the value obtained with the radio maps by \citet{dickey90} does not resolve the small scale, $<$1$^{\prime}$, non-uniformity of $N_H$ and 
does not include the possible contribution of molecular hydrogen, probably underestimating the true value. Indeed also the 
second source in the \chandra~field of view, AX~J1745.6$-$2901 located  $\sim$18$^{\prime}$ from our target source, is heavily obscured with $N_H = (34\pm2) \times 10^{22}\,{\rm cm^{-2}}$ \citep[average Galactic value in the source direction $\sim$1.2$\times10^{22}$\,cm$^{-2}$][]{dickey90} possibly related to the patchy nature of $N_H$ towards the Galactic Center.

Our NIR observations showed that 2MASS~J17452768$-$2919534 underwent a slight brightening in the $J$-band during the X--ray outburst, while it remained constant in the  $H$ and $K_s$ bands. This variability detected in the bluer bands rather than in the $K_s$ band could be consistent with an enhanced emission from an accretion disk of an LMXB,  in which the donor star (or even the jet emission) dominates the
NIR flux in $H$ and $K_s$, with an increasing disk contribution at bluer wavelengths \citep[e.g.][]{charles06, khargharia10}.

To estimate the extinction towards the source, we modify the relationship of \citet{predehl95} to account for the fact that the absorption model of \citet{wilms00} used throughout  the paper fits neutral columns $\approx$30\% larger than the model used by \citet{predehl95}, hence we assume A$_{V}\sim {\rm N_H}/$2.7$\times 10^{21}\,{\rm cm}^{-2}$ \citep[as in][]{nowak12}.

Using  the ratio A$_{K_s}$/A$_{V}$=0.112 and A$_{J}$/A$_{V}$=0.282 \citep{rieke85}, our broad-band observed column density 
${\rm N_H}\sim$11.9$\times 10^{22}\,{\rm cm}^{-2}$  translates into an absolute $K_s$ magnitude $M_{K_s}$=-8.1\,mag with an assumed distance of 8\,kpc ($M_{K_s}$=-7.1\,mag at 5\,kpc), an absolute $J$ magnitude $M_{J}$=-10.9\,mag ($M_{J}$=-9.9\,mag at 5\,kpc) and $M_{J}$-$M_{K_s}$=-2.8\,mag.
The obtained $K_s$ band value is compatible with an M-type companion in the case of a red giant, however the $M_{J}$-$M_{K_s}$ value does not seem to fit with any spectral type \citep[see Figure~1 in][where -0.5$<M_{J}$-$M_{K_s}<+$1.5]{chaty02}. Notwithstanding the patchy nature of $N_H$ towards the Galactic Center, it would take an unusually low extinction to match the obtained value to a given spectral type. 

We note that similar results are obtained also if we use more recent interstellar extinction laws, such as by \citet{nishiyama09} and \citet{guver09}. In both cases, the $K_s$-band indicates an M-type red giant companion (LMXB), whereas  $M_{J}$-$M_{K_s}$ does not belong to any spectral type ($M_{J}$-$M_{K_s}$=-3.2\,mag and -4.45\,mag, respectively). 

 On the other extreme, we could consider that all the absorbing material is local to the accreting compact object alone,  while the companion is not enshrouded in it. In this case, as in some \integral~highly absorbed HMXBs \citep[e.g.,][and references therein]{chaty13}, we estimate the extinction towards the source   only using the expected line of sight absorption  \citep[$\sim$1.2$\times10^{22}$\,cm$^{-2}$][]{dickey90}. This leads to an absolute $K_s$ magnitude \mbox{$M_{K_s}=$-3.9\,mag} for an assumed distance of 8\,kpc (\mbox{$M_{K_s}$=-2.9\,mag} at 5\,kpc), compatible with a K/M-type red giant companion (LMXB) or a B-type main sequence star (HMXB). Similarly to the previous case, the $M_{J}$-$M_{K_s}$ value obtained \mbox{($M_{J}$-$M_{K_s}$=$+$3.6\,mag)} does not seem to match  any spectral type. Most likely, \mysou~lies somewhere in between, with \textit{some} of the absorbing material local to the accreting compact object. Furthermore, part of the $M_{J}$-$M_{K_s}$ discrepancy could mean that we are seeing the contribution from the accretion disk (suggesting an LMXB). Indeed, NIR increases up to  $\sim$4--7\,mag have been seen in the high-soft state of LMXBs \citep{charles06}, and of about 3\,mag in the low-hard state \citep{chaty03}. 

Since \mysou~is right in the crowded region of the Galactic center, it is possible that the associated NIR counterpart discussed up to now is not the correct one, with the real one lying behind our candidate, or within a blend with other bright stars dominating the NIR scene. As a first order estimate however, we note that in Figure~\ref{fig:Kband} we have about 25 comparably bright sources to \mysou~in an area about 790 times that of the \chandra~error circle, resulting in a low chance coincidence probability, of about 3\%.

For completeness, we investigate also the active galactic nucleus (AGN) possibility. There are a few examples of AGNs located in the Galactic plane, and even towards the Galactic center \citep{chaty08, zurita09,tomsick12}. However, no extragalactic object is 
known to be located within a radius of 30$^{\prime}$ from the Galactic Centre (SIMBAD), with \mysou~being less than 24$^{\prime}$ from it. Furthermore, \mysou, with its $\sim$2$\mathrm{\times 10^{-10}~erg~cm^{-2}~s^{-1}}$
 in 20--40\,keV (see Figure~\ref{fig:allLCR}) would place itself among the brightest AGNs detected by \integral. Comparing with \citet{beckmann09}, we see that out of 199 AGNs detected  with \integral~above 20\,keV, only four do have a flux brighter than 1$\mathrm{\times 10^{-10}~erg~cm^{-2}~s^{-1}}$ in the 20--40\,keV: Mrk 421 (BLLac), NGC~4151 (Sy1.5), Cen~A (Sy 2) and the Circinus Galaxy (Sy 2). The latter three are Seyfert galaxies and, similarly to \mysou,  are highly absorbed in X--rays, but while in their case this results in a significant infrared $K_s$  emission due to the thermal radiation from the dust \citep[$\sim$4\,mag for Cen~A and $\sim$7.5\,mag for NGC~4151,][]{skrutskie06}, in our case we reach a level of $\sim$11\,mag, much dimmer. On the other hand, this source seems too absorbed to be a blazar AGN (e.g. for Mrk~421, $\rm N_H$=0.08$\times 10^{22}\,{\rm cm}^{-2}$ is obtained). Finally, the AGN hypothesis is discarded also by the NIR images since if the X--rays were from an AGN similar to the above, the source would be nearby and thus we would observe a NIR extended source \citep[several arcsecs to even arcminutes, see e.g. the dimmer IGR~J09026--4812 with its 4.9$^{\prime\prime}$ semi-major axis in $K_S$,][]{zurita09}. This is clearly not the case in the archival and GROND NIR maps.   \\

\section{Summary}

On 2014 September 27th, \integral~discovered the new transient \mysou, 24$^{\prime}$ away from the Galactic
center.  The outburst lasted at least about 60 days, with a longer monitoring being hampered by solar constraints.
We studied the long-term X--ray light-curve and
broad-band spectra of \mysou~using data from three high energy missions: \chandra, \swift~and \integral. 

The outburst X--ray light-curve shows a double peak with a clear $\sim$4--5\,d flux decrease seen
in \emph{all} X--ray bands. This may suggest a mass transfer change from the companion in
an eccentric orbit, or a change
of the absorbing medium (e.g. dips from intervening matter, eclipses by
the donor or by a tilted and/or warped accretion disk).

The outburst peak flux  is
$\sim$5.5$\mathrm{\times 10^{-10}~erg~cm^{-2}~s^{-1}}$ (15--85\,keV) corresponding to 0.5--100\,keV peak luminosity of 9$\mathrm{\times 10^{36}~erg~s^{-1}}$ at 8\,kpc, in the range of X--ray binary outbursts. The broad band spectra of the source are compatible with an absorbed power-law ($\Gamma\sim$1.6-1.8 and ${\rm N_H}\sim$10--12$\times 10^{22}\,{\rm cm}^{-2}$), with no trace of a cut-off in the data up to about 100\,keV. This is compatible with an LMXB in the low-hard state.  The detection of a broad iron line by \nustar~\citep{tendulkar14} strengthens this association. Up to the time of writing, there is no indication  on the nature of the compact object (i.e. pulsations or type-I X-ray bursts that would point to a NS).

With \chandra, we  determined the most accurate  X-ray position of \mysou, enabling a candidate NIR counterpart search in the crowded 
field towards the Galactic center. The X--ray position of  \mysou~is compatible with the NIR  source 2MASS~J17452768--2919534. 
Archival (2MASS, UKIDSS, VVV) and new  (GROND) NIR observations of the source taken during the 2014 outburst have been investigated. 

The obtained $K_s$ band values are compatible with an K/M-type companion in the case of a red giant with 
a hint of brightening in the $J$ band. This is consistent with an enhanced emission from the accretion disk of an LMXB.
Moreover, the $M_{J}$-$M_{K_s}$ value does not seem to match any spectral type and this could mean that  
we are seeing the contribution from the accretion disk, with some of the absorbing material possibly local to the accreting object alone.

It is not straightforward to unveil the nature of this elusive source. However, considering the outburst  X--ray properties  
and the NIR ones, \mysou~is most likely an LMXB, with no current indication on the nature of the compact object.

\acknowledgments 
We thank the \chandra~team for their rapid response in
scheduling and delivering the observation.\\
This work is partly based on observations with \integral, an ESA project
with instruments and science data center funded by ESA member states, Czech Republic and Poland, and with the
participation of Russia and the USA. \\
This research has made use of the \integral~sources page http://irfu.cea.fr/Sap/IGR-Sources/.\\
Part of the GROND funding (both hardware and personnel) was generously granted from the Leibniz-Prize to Prof. G. Hasinger, Deutsche
Forschungsgemeinschaft (DFG) grant HA 1850/28-1. Sebastian Schmidl acknowledges support by the Th\"uringer Ministerium f\"ur Bildung, Wissenschaft
und Kultur under FKZ 12010-514.  Based on data products from observations made with ESO Telescopes at the La Silla or Paranal Observatories
under ESO programme ID 179.B-2002.\\
This publication makes use of data obtained as part of the UKIRT Infrared Deep Sky Survey and of data products from the Two Micron All Sky Survey, which is a joint project of the University of Massachusetts and the Infrared Processing and Analysis Center/California Institute of Technology, funded by the National Aeronautics and Space Administration and the National Science Foundation.

AP acknowledges the Italian Space Agency financial support
INTEGRAL ASI/INAF agreement no. 2013-025.R.0. MAN acknowledges support from NASA Grant G04-15027X. JC acknowledges financial support from ESA/PRODEX Nr. 90057.
J.R. acknowledges funding supports from the UnivEarthS Labex program of Sorbonne Paris Cit\'e (ANR-10-LABX-0023 and ANR-11-IDEX-0005-02).
 AP thanks Lara Sidoli and Volker Beckmann for useful discussions. We thank the anonymous Referee for the accurate reading of the manuscript.

%\bibliographystyle{apj}
%\bibliography{biblio}

\end{document}